\documentstyle[12pt,colordvi,epsfig]{article}
\begin{document}
\centerline{\Large\bf Evidence for the presence of dust in}
\centerline{\Large\bf intervening QSO absorbers from the}
\centerline{\Large\bf Sloan Digital Sky Survey}
\vspace{0.4in}
\noindent P. Khare\footnote{Utkal University, India
\hspace{1.27in}$^2$Univ. Chicago, USA}, D. G. York$^2$, D. Vanden
Berk\footnote[3]{Pennsylvania State University, USA\hspace{0.53in}
$^4$Univ. South Carolina, USA}, V. P. Kulkarni$^4$, A. P. S.
Crotts\footnote[5]{Columbia University, USA
\hspace{1.05in}$^6$Northwestern University, USA},\break D. E. Welty$^2$,
J. T. Lauroesch$^6$, G. T. Richards\footnote[7]{Princeton University
Observatory, USA \hspace{0.32in}$^8$New York University, USA}, Y.
Alsayyad$^2$, A. Kumar$^8$, B. Lundgren\footnote[9]{Univ. Illinois,
Urbana Champaign, USA\hspace{0.3in}$^{10}$Univ. California, Santa
Barbara, USA}, N. Shanidze$^2$, J. Vanlandingham$^2$, B. Baugher$^{10}$,
P. B. Hall\footnote[11]{York University,
Canada \hspace{1.25in}$^{12}$Inst. for Adv. Study, Princeton, USA}, E.
B. Jenkins$^7$, B. Menard$^{12}$, S. Rao\footnote[13]{Univ. Pittsburgh,
USA}, D. Turnshek$^{13}$, C.W. Yip$^{13}$\\

\begin{abstract}
We find evidence for dust in the intervening QSO absorbers from the
spectra of QSOs in the Sloan Digital Sky Survey Data Release 1. No
evidence is found for the 2175 {\AA} feature which is present in the
Milky Way dust extinction curve. The extinction curve resembles the
SMC extinction curve. The observed $\Delta$(g-i) excess for QSOs with
strong absorption systems appears to be a result of the reddening due to
dust in the intervening absorbers.\\

\end{abstract}

\section{Introduction} Dust has long been suspected to be a constituent
of QSO absorbers. The presence of dust in DLAs is suggested by the
differential depletion of the refractory elements like Cr and Fe with
respect to the volatile elements like Zn (Khare et al. 2004). [Cr/Zn] has
been found to be correlated with E(B-V) in a sample of a few Sloan
Digital Sky Survey (SDSS) QSOs (Khare et al. 2004). Yet another signature
of the Milky Way type of dust, namely the 2175 {\AA} feature in the
extinction curve has been detected in a few individual QSOs (e.g. Wang et
al. 2004). Malhotra (1997) reported the detection of the 2175 {\AA}
feature in the composite spectra of 96 QSOs. A much larger, homogeneous
sample of QSOs is now available thanks to the SDSS (York et al. 2000).
Here we study the SDSS Data Release 1 (DR1; Abazajian et al. 2003,
Richards et al. 2003) having over 16,000 QSOs, for presence of dust in
the intervening absorbers. 
\section{Sample definition} The absorption system identification has been
done by the pipeline developed by York et al. (2005)05. To generate our
``absorber" sample we collected all absorption systems with 4 or more
detected lines outside the Lyman alpha forest, in non-BAL QSOs, and with
(i) redshift between 1.0 and 1.9 (to ensure the observability of the 2175
{\AA} feature) (ii) W$^{rest}_{MgII2796}\; >$ 0.3 {\AA} and (iii)
relative velocity w.r.t. the QSO $\beta$ $>$ 0.01 (to ensure the
intervening nature of the absorbers). The final sample consisted of 815
absorbers. We also constructed a ``non-absorber" sample consisting of DR1
QSOs having no absorption systems in their spectra by matching z$_{em}$
and i magnitude on a one to one basis between the ``absorber" and
``non-absorber" samples. 
\section{Composite spectra and evidence for reddening} The spectra of
QSOs in the ``absorber" sample were shifted to the absorber rest frame.
The same shifts were applied to the spectra of the corresponding
``non-absorber" QSOs. Geometric mean composite spectra for the two
samples were generated. Fig. 1 shows the resulting spectra for the two
samples and their ratio along with the best fit SMC extinction curve and
the best fit Milky Way extinction curve for $\lambda >$ 1500 {\AA}. The
presence of reddening in the ``absorber" sample is clearly indicated. The
2175 {\AA} feature is definitely absent. We obtain a mean E(B-V)=0.013
$\pm{0.002}$ using the SMC extinction curve which seems to fit the
spectra better. 

\begin{figure}[h]
\begin{minipage}[t]{6.5cm}
\begin{flushleft}
\includegraphics[width=7.0cm,clip]{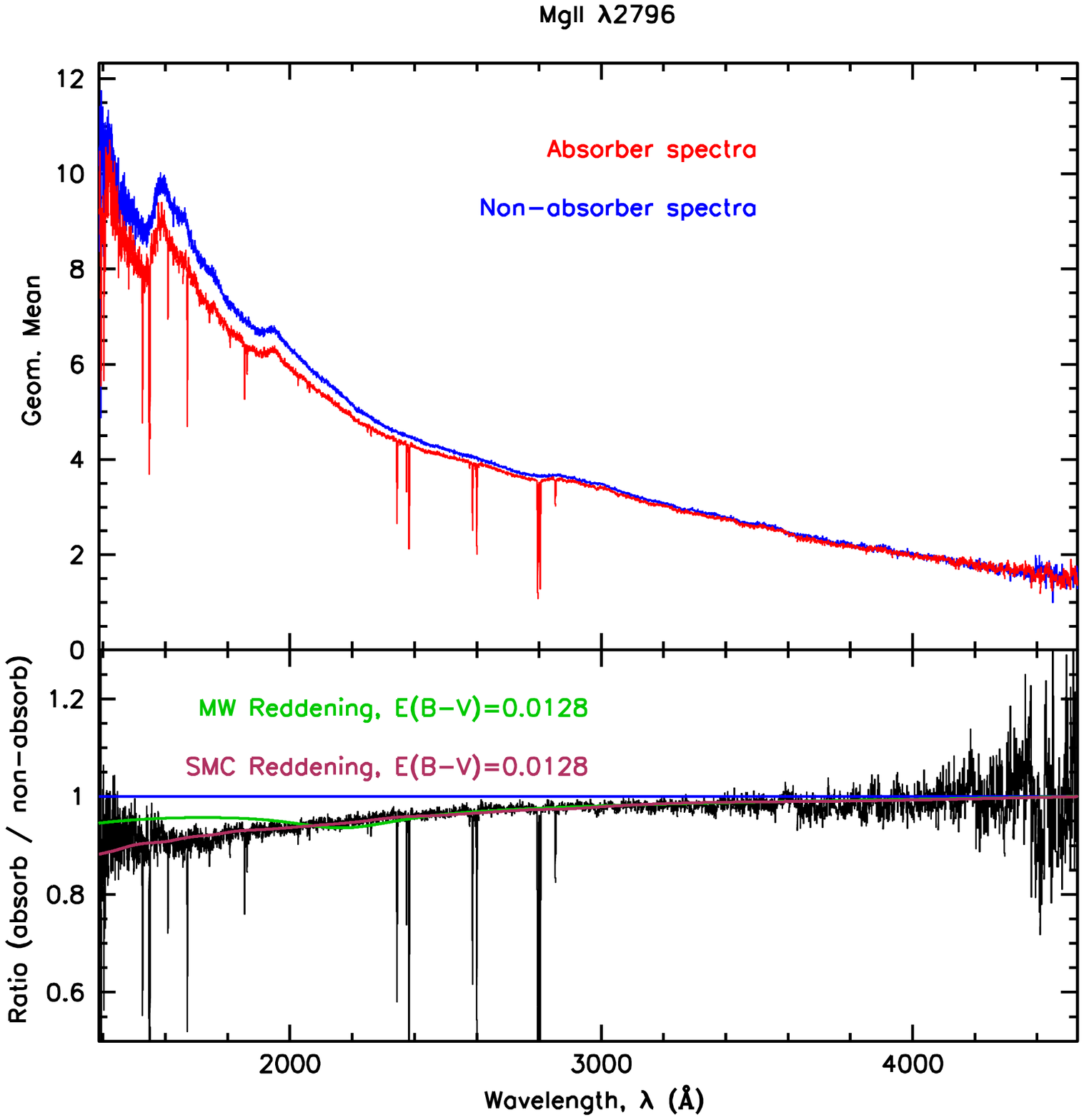}
\caption{The upper panel shows the composite spectra in the absorber rest
frame, of the ``absorber" and the ``non-absorber" samples. The lower
panel shows the ratio of the two composite spectra and the best fit Milky
Way and SMC extinction curves.}
\end{flushleft}
\end{minipage}
\hfill
\begin{minipage}[t]{6.5cm}
\begin{flushleft}
\includegraphics[width=7.0cm,clip]{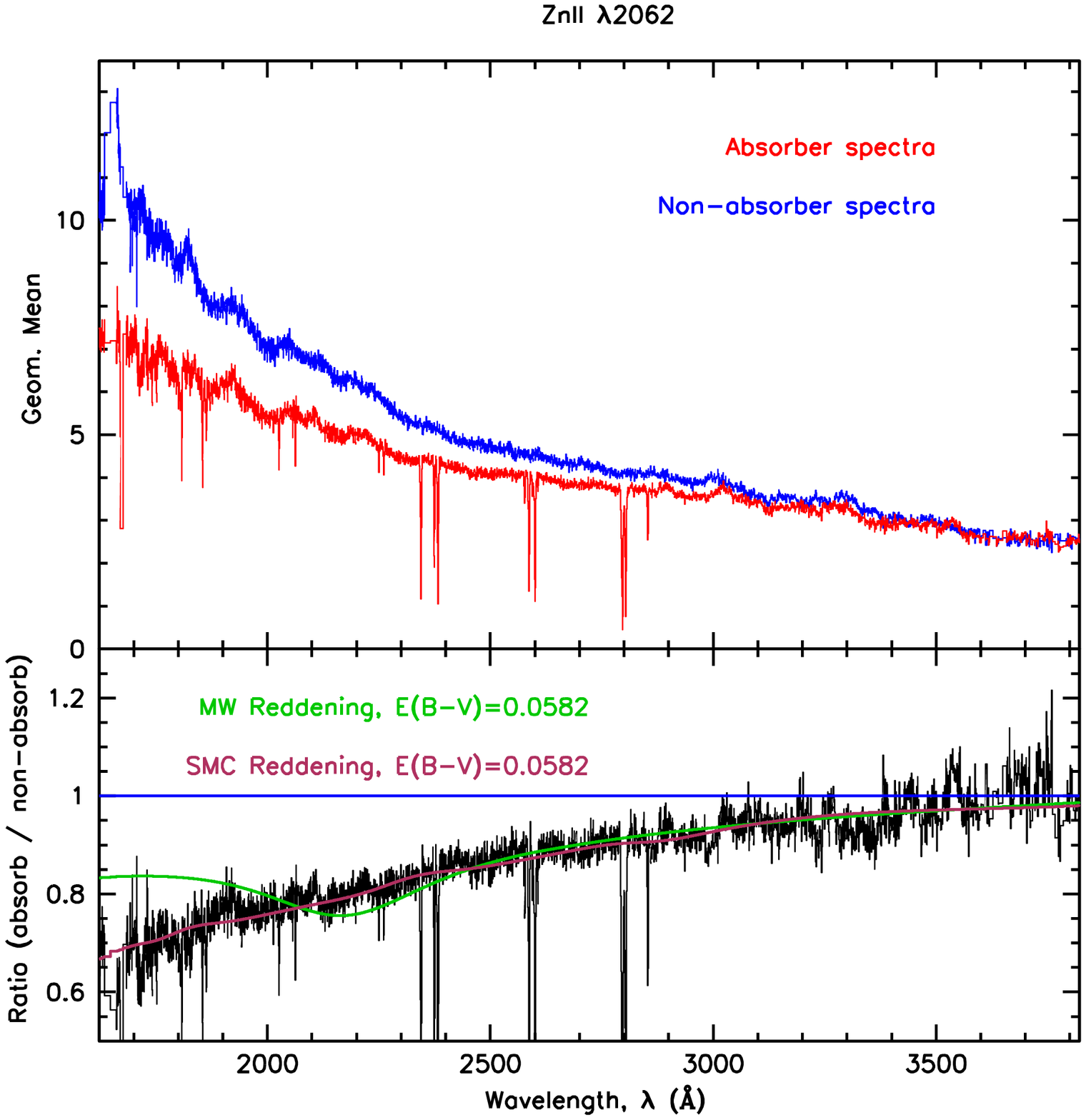}
\caption{The upper panel shows the composite spectra in the absorber rest
frame, of systems in which the Zn II/Cr II $\lambda$2062 blend has been
observed, for the ``absorber" sample and the corresponding
``non-absorber" sample. The lower panel shows the ratio of the two
composite spectra and the best fit Milky Way and SMC extinction curves.}
\end{flushleft}
\end{minipage}
\end{figure}

\begin{figure}[ht]
\begin{minipage}[t]{6.5cm}
\begin{flushleft}
\includegraphics[width=7.0cm,clip]{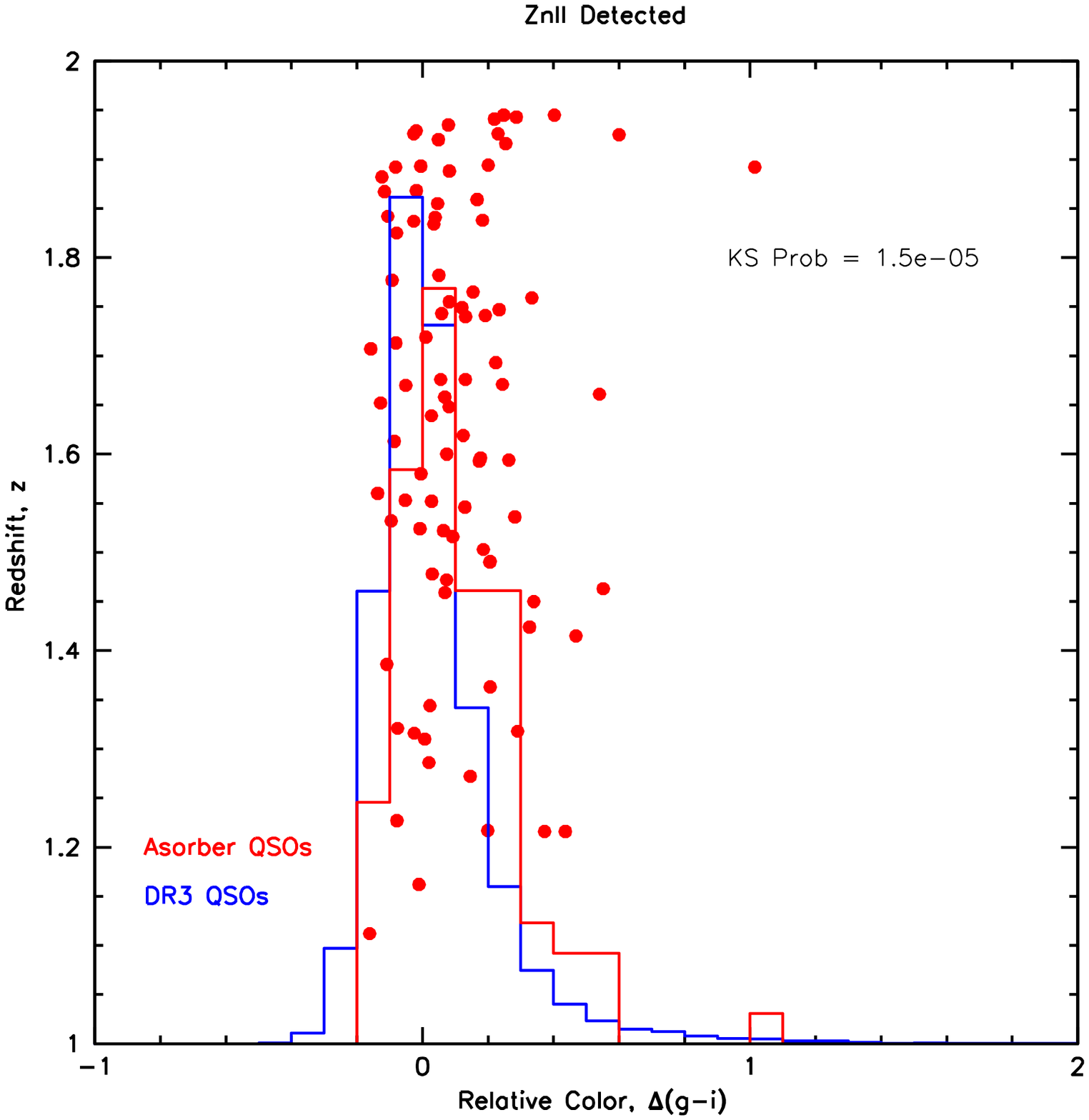}
\caption{Histogram of the colour excess $\Delta$(g-i) for the DR3 
and the sample of systems in which the Zn II/Cr II $\lambda$ 2062 blend
was detected. Values of $\Delta$(g-i) for these systems have
been plotted against their redshifts (dots). The KS probability
that the two samples are drawn from the same populations is indicated.}
\end{flushleft}
\end{minipage}
\hfill
\begin{minipage}[t]{6.5cm}
\begin{flushleft}
\includegraphics[width=7.0cm,clip]{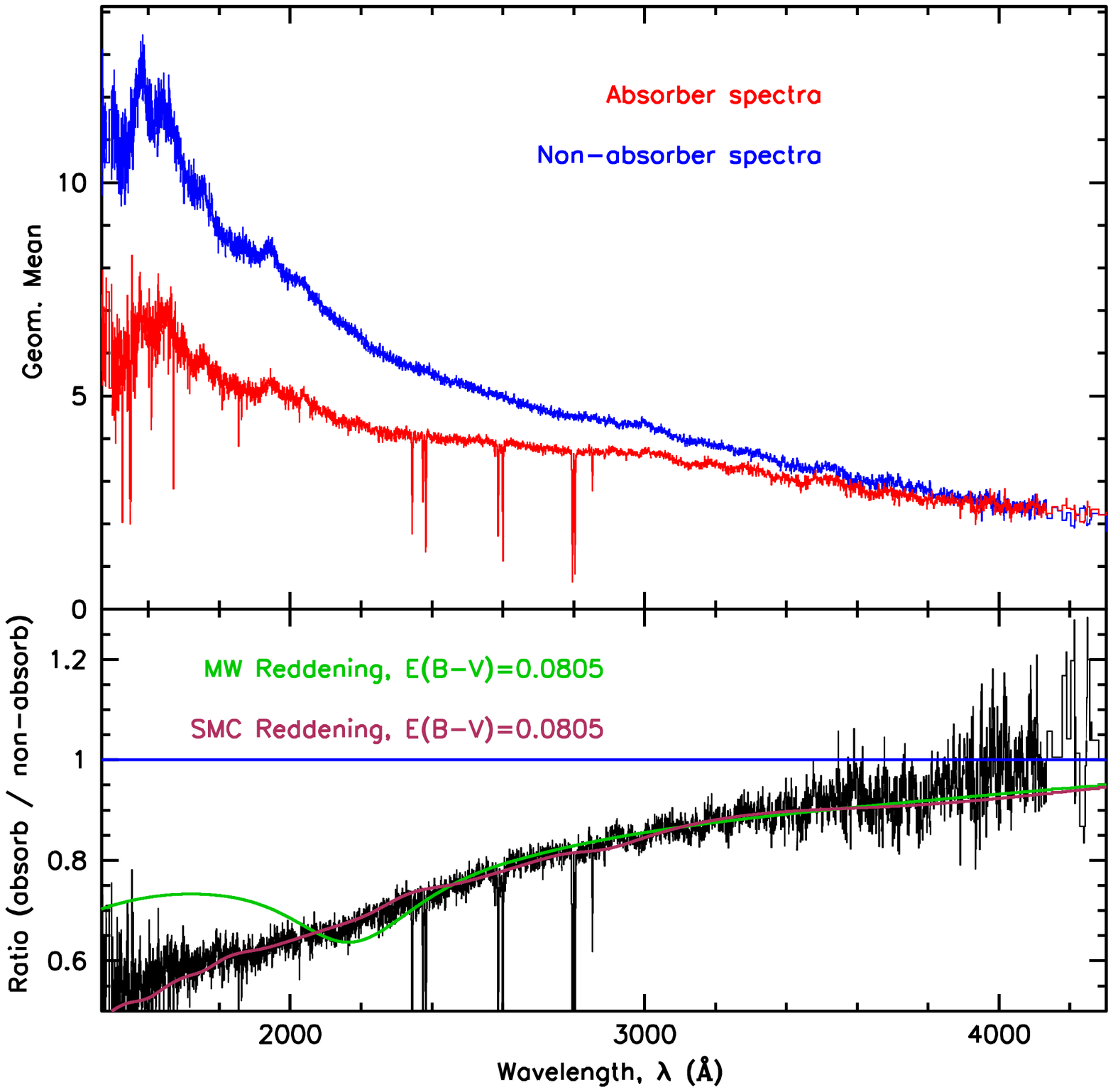}
\caption{The upper panel shows the composite spectra in the absorber rest
frame, of systems with $\Delta$(g-i) $>$ 0.2, for the ``absorber" sample
and the corresponding ``non-absorber" sample. The lower panel shows the
ratio of the two composite spectra and the best fit Milky Way and SMC
extinction curves.}
\end{flushleft}
\end{minipage}
\end{figure}

To study the dependence of reddening on the absorber properties, we have
generated composite spectra for several ``absorber" and ``non-absorber"
subsamples based on the strength of absorption lines of various species,
on absorber redshifts, on $\beta$ and on the i magnitudes. The E(B-V)
values are found to be insensitive to $\beta$ and i magnitudes. We find
that E(B-V) is high for systems with strong Mg II $\lambda$2796, while
systems with weak Mg II $\lambda$2796 show insignificant reddening. The
E(B-V) is highest (0.058) for the sample of systems in which the Zn II/Cr
II $\lambda$2062 blend was detected (Fig. 2).
\section{Colour excess} The colour excess of a QSO w.r.t. the mean colour
($\Delta$(g-i); the difference between the observed (g-i) for a given QSO
and the median (g-i) for all QSOs at that redshift (Richards et al.
2003)) is indicative of the extinction in its spectrum. Fig. 3 shows
histograms for $\Delta$(g-i) for the subsample of absorbers in which Zn
II/Cr II $\lambda 2062$ blend has been detected and for all the non-BAL
QSOs in the Data Release 3 (DR3; Abazajian 2005). The ``absorber"
subsample is significantly redder than average DR3 QSOs. Thus strong line
absorbers appear to significantly redden QSOs. Fig. 4 shows the composite
spectra and the fitted extinction curves for the subsample of QSOs with
$\Delta$(g-i) $>$ 0.2. A high value of E(B-V) (0.081) is obtained for
this sample. Note that the E(B-V) values are likely to be overestimates
due to the presence of multiple absorbers along some lines of sight and
due to the fact that the absorber sample may not be complete due to the
presence of low S/N spectra in the sample.

\section{Conclusions}
\begin{itemize}
\item  The intervening absorbers in the QSO spectra redden the 
spectra.
\item  The extinction curve due to the intervening absorbers is similar
to the SMC extinction curve.
\item  No evidence is found for the presence of the 2175 {\AA}
feature.
\item  The observed $\Delta$(g-i) excess for QSOs with strong absorption
systems appears to be a result of the reddening due to dust in the
intervening absorbers.
\end{itemize}

Funding for the creation and distribution of the SDSS Archive has 
been provided by the Alfred P. Sloan Foundation, the Participating 
Institutions, the National Aeronautics and Space Administration, the 
National Science Foundation, the U.S. Department of Energy, the 
Japanese Monbukagakusho, and the Max Planck Society. The SDSS Web 
site is http://www.sdss.org/\\

\end{document}